\documentstyle[psfig,preprint,aps]{revtex}

\tightenlines

\def\beq{\begin{equation}}
\def\eeq{\end{equation}}
\def\bea{\begin{eqnarray}}
\def\eea{\end{eqnarray}}

\begin{document}
\hoffset-1cm
\draft
\title{Bremsstrahlung from an Equilibrating Quark-Gluon Plasma
\footnote{Supported by BMBF, GSI Darmstadt, DFG, and Humboldt foundation}}

\author{Munshi G. Mustafa$^1$\footnote{Humboldt fellow and on leave of
absence from Saha Institute of Nuclear Physics, 1/AF Bidhan Nagar,
Calcutta 700 064, India}
 and Markus H. Thoma$^2$\footnote{Heisenberg fellow}}
\address{$^1$Institut f\"ur Theoretische Physik, Universit\"at Giessen, 35392
Giessen, Germany}
\address{$^2$Theory Division, CERN, CH-1211 Geneva 23, Switzerland}

\date{\today}

\maketitle

\begin{abstract}
The photon production rate from a chemically equilibrating quark-gluon plasma 
likely to be produced at RHIC (BNL) and LHC (CERN) energies is estimated
taking into account bremsstrahlung.
The plasma is assumed to be in local thermal equilibrium, but with a phase 
space distribution that deviates from the Fermi or Bose distribution by space-time 
dependent factors (fugacities). The photon spectrum is obtained by integrating the 
photon rate over the space-time history of the plasma, adopting a boost invariant 
cylindrically symmetric transverse expansion of the system 
with different nuclear profile functions. Initial 
conditions obtained from a self-screened parton cascade calculation and, for comparison, 
from the HIJING model are used. Compared to an equilibrated plasma at the same 
initial energy density, taken from the self-screened parton cascade,
a moderate suppression 
of the photon yield by a factor of one to five depending on the collision
energy and the photon momentum is observed. The individual contributions
to the photon production, however, are completely different in the
both scenarios.
\end{abstract}

\bigskip

\medskip


\narrowtext
\newpage

\section{Introduction}
In ultrarelativistic heavy ion collisions photons are emitted during the 
entire life time of the fireball. Since their mean free path is large
compared to the size of the fireball \cite{Tho95}, energetic photons escape 
from it without any interaction. Therefore photons with large transverse 
momenta probe the early, hot stage of the collision and may serve as a direct
signature for the quark-gluon plasma (QGP) formation at RHIC and LHC
\cite{Ruu92}. In order to use photons as a probe for the QGP, we have to 
predict the thermal spectrum from the QGP as well as from the hadronic phase.
For this purpose we have to convolute the photon production rates from both 
phases with the space-time evolution of the fireball. Here we want to focus 
on a QGP phase possibly created at RHIC and LHC energies. Recent calculations
of the photon production rate from the QGP, taking into account bremsstrahlung,
resulted in a significantly larger rate compared to previous investigations
\cite{Aur98}. These rates have been implemented in a hydrodynamical
calculation assuming a local thermal and chemical equilibrium 
\cite{Sri99,Ste99}.
However, at RHIC and LHC the QGP phase is not expected to be in chemical
equilibrium, {\it i.e.}, the phase space of quarks and gluons will be
undersaturated probably \cite{Bir93,Xio94,Wan97}. Using non-equilibrium
photon rates without bremsstrahlung and generalizing the hydrodynamical 
calculations to chemical non-equilibrium, the photon spectrum from a 
chemically non-equilibrated QGP has been estimated \cite{Str94,Tra96,Sri97}. 
Due to the undersaturation in particular of the quark 
component a suppression of the photon yield has been observed \cite{Tra96}.
In order to make up-to-date predictions for the thermal photon spectrum
of a QGP at RHIC and LHC, the both competing effects of the enhanced photon 
rate considering bremsstrahlung and the reduction of the photon yield due to 
the chemical non-equilibrium should be combined. This is the aim of the 
present investigation.

In the next section we present an estimate of the non-equilibrium photon 
rates taking into account bremsstrahlung. In Sec.III we summarize briefly
our hydrodynamical description, before we will discuss our results in Sec.IV 
and draw our conclusions in Sec.V.

\section{PHOTON PRODUCTION FROM QUARK-GLUON PLASMA}

The production rate of energetic real photons in an equilibrated QGP has been 
calculated using the Hard Thermal Loop (HTL) resummation technique
\cite{Kap91,Bai92}. Here only the lowest order contributions, namely
quark-antiquark annihilation and Compton scattering (corresponding to a one-loop
polarization tensor containing a HTL resummed quark propagator in the case
of a soft quark momentum), has been taken into 
account. An estimate for these contributions in chemical non-equilibrium
has been given by computing these rates from the tree level scattering 
diagrams and using off-equilibrium distribution functions for the
external quarks and gluons \cite{Tra96}. For this purpose the equilibrium 
distributions have been multiplied by space-time dependent
fugacity factors $\lambda_{g,q}$ describing
the deviation from equilibrium according to Ref.\cite{Bir93,Sri97a}.
Also the infrared cutoff in these rates given by the effective in-medium 
quark mass has been generalized to non-equilibrium. Neglecting Pauli
blocking and Bose enhancement effects, which is justified since the
average momenta of the partons is $3T$ in the baryon free QGP,
the following result for the rates due to Compton scattering and 
quark-antiquark annihilation has been obtained \cite{Str94,Sri97},
\begin{equation}
\left. E\frac{{\rm d}N_\gamma}{{\rm d}^4x\,{\rm d}^3p}\right 
|_{\sf com.}^{\sf ceq.}= 
\frac{2\alpha\alpha_s}{\pi^4}
\lambda_q \lambda_g T^2 \left ( \sum_f e_f^2\right ) e^{-E/T}
\left[ \ln\left(\frac{4ET}{\kappa_c^2}\right)
+\frac{1}{2}-C\right],
\label{comp}
\end{equation}
and 
\begin{equation}
E\left.\frac{{\rm d}N_\gamma}{{\rm d}^4x\,{\rm d}^3p}\right 
|_{\sf ann.}^{\sf ceq.}
= \frac{2\alpha\alpha_s}{\pi^4} \lambda_q \lambda_{\bar{q}}
T^2 \left ( \sum_f e_f^2\right )e^{-E/T}
\left[ \ln\left(\frac{4ET}{\kappa_c^2}\right)
-1-C\right].
\label{ann}
\end{equation}
Here $C=$ 0.577216\ldots,  and
$\kappa_c =2m_q^2$, where the thermal mass of the quarks in the
non-equilibrated medium is given as
\begin{equation}
m_q^2 = \frac{4\pi \alpha_s }{9} 
\left (\lambda_g + \frac{\lambda_q}{2}\right ) T^2.
\label{qmass}
\end{equation}
Alternatively non-equilibrium rates can be calculated by generalizing the
HTL technique to quasistatic chemical non-equilibrium situations
\cite{Bai97,Car99}. In the case of the photon production rate this more
consistent, but also more elaborate, method leads to quantitatively
very similar results as the simplification used above \cite{Bai97}.

Recently Aurenche et al.\cite{Aur98} have shown that there are additional
contributions to the thermal photon rate at the same order $\alpha_s$
coming from two-loop diagrams within the HTL method. The corresponding
physical processes are bremsstrahlung and quark-antiquark annihilation, where
the quark (or antiquark) scatters of a parton from the QGP. The latter process
becomes important especially for energetic photons as the rate is proportional
to $ET$, whereas the rates from bremsstrahlung, annihilation without 
rescattering, and Compton scattering are proportional to $T^2$. Although
the Compton and annihilation contributions are enhanced in the weak coupling limit
by a logarithmic
factor $\ln (1/\alpha_s)$ compared to the bremsstrahlung processes, the latter
ones dominate for realistic values of the coupling constant by a factor of
5 or more \cite{Aur98,Sri99}. Of course, the extrapolation of these 
perturbatively 
calculated rates to realistic values of the coupling constant, for which
we will use $\alpha_s =0.3$ in the following, is questionable.
However, in view of the lack of other methods for calculating dynamical 
quantities such as production rates in thermal field theory so far, we assume that
these results can be used as a rough estimate at least in the case of high
energy photons ($E\gg T$), in which we are interested.

For estimating the bremsstrahlung contributions to the non-equilibrium
rate we adopt the same approximations as in (\ref{comp}) and (\ref{ann}).
This means that we start from the scattering diagrams corresponding
to these processes and ascribe non-equilibrium distributions to the external 
partons in the entrance channels. Restricting ourselves to $t$-channel 
diagrams, which dominate because the exchanged gluon is soft \cite{Aur98}, 
and considering the different spin, color, and
flavor ($N_f=2$) factors, we find
for the bremsstrahlung process 
\begin{equation}
E\left.\frac{{\rm d}N_\gamma}{{\rm d}^4x\,{\rm d}^3p}\right 
|_{\sf brem.}^{\sf ceq.}
=\frac{2N_cC_F}{\pi^5}\alpha\alpha_s 
\left (\frac{4}{7} \lambda_q^2 + \frac{3}{7}\lambda_g\lambda_q\right )
\left ( \sum_f e_f^2\right ) T^2 e^{-E/T} (J_T-J_L) \ln (2)\ , 
\label{brem2lc}
\end{equation}
and for the annihilation with scattering process 
\begin{equation}
 E\left.\frac{{\rm d}N_\gamma}{{\rm d}^4x\,{\rm d}^3p}\right 
|_{\sf aws}^{\sf ceq.}
=\frac{2N_cC_F}{3\pi^5}\alpha\alpha_s 
\left (\frac{2}{5} \lambda_q^3 + \frac{3}{5}\lambda_g\lambda_q^2\right )
\left ( \sum_f e_f^2\right ) ET e^{-E/T} (J_T-J_L)  \ \ . 
\label{awsc}
\end{equation}
Here the constants $J_T\simeq 4.45$ and $J_L\simeq -4.26$ for two flavors
are the same as in equilibrium, since their dependence on the fugacities can 
be neglected. For they are functions of $m_q/m_g$ only and the square of the 
effective quark mass $m_q^2$, Eq.(\ref{qmass}), and of the effective gluon 
mass $m_g^2$ are proportional to the gluon fugacity, if the much smaller 
quark fugacity is neglected.
It should be noted that in Eq.(\ref{awsc}) the combination of the different 
parton fugacities appears in cubic power, instead of quadratically as for
the other processes, 
due to the fact that this particular process involves three particles in the 
entrance channel. 

\section{\bf HYDRODYNAMIC EXPANSION AND CHEMICAL EQUILIBRATION}

To evaluate the thermal photon spectrum one has to convolute these 
emission rates with the space time history of the expanding fireball
out of chemical equilibrium. We do not want to repeat here the
calculations of the chemical evolution of the QGP by means of rate equations
\cite{Bir93} and its implementation in a hydrodynamical code, but refer
the reader to Ref.\cite{Sri97,Sri97a}. In this hydrodynamical model, which we are
adopting here, the transverse expansion of the QGP has been taken into 
account.

The essential input needed are the initial
conditions for the fugacities and temperature at the time at which local 
thermal equilibrium is achieved. In order to take into account the 
uncertainties in these initial conditions we consider various possibilities
predicted by two microscopic models for ultrarelativistic heavy ion collisions, 
namely SSPC \cite{Esk96} and HIJING \cite{Wan97}. The initial
conditions we are using here are tabulated in Tab.I. The ones
denoted by (I) are the original HIJING predictions, while the set (II) 
is obtained by multiplying the original initial fugacities by a factor of 4
and by decreasing the initial temperature somewhat, in order to
take into account possible uncertainties in the model such as the neglect of
soft parton production from the color field \cite{Sri97}.

In addition to the different initial conditions, we also need a nuclear 
profile function for the fireball to solve the hydrodynamical equations 
numerically as matter distributions with sharp edges are difficult to 
use in numerical simulations \cite{Ger86}. Another aim of our investigation 
is to study the dependence of the photon spectra on different choices of
the profile function. For this purpose we consider in the
following two profile functions. The first one is a Fermi-like profile 
function \cite{Ger86}
\begin{equation}
T_A(r) = \frac{1}{e^{\left (r - R_T \right )/\delta}+1} \ \ , 
\label{fermi}
\end{equation}
where $r$ is the transverse coordinate, $R_T$ is the transverse radius of 
the nucleus and $\delta$ is the surface thickness, which has also been used 
in Ref.\cite{Sri97}. 
The second one is the wounded-nucleon profile \cite{Mue97} given by
\begin{equation}
T_A(r) = \frac{3}{2} \sqrt {1 - \frac{r^2}{R_T^2}} \  \ .
\label{wound}
\end{equation}
Since the initial conditions shown in Tab.I describe averages over the 
transverse cross section, $\pi R_T^2$, of the colliding nuclei 
in a central collision, one needs to modify them for different nuclear profiles.
This implies that the system will have higher initial energy density and 
fugacities using the wounded-nucleon profiles, but also the preexistence 
of a density gradient over the whole transverse area resulting in a faster 
onset of the transverse expansion throughout the plasma volume \cite{Mue97}. 

\section{RESULTS}

In Fig.1 we present the thermal photon production from the quark phase as a function 
of its transverse momentum at RHIC energies with SSPC initial conditions and the 
Fermi-like profile function both in the equilibrium and equilibrating scenario,
where we assumed the same initial energy densities in both cases given in Tab.I.
The initial temperature in the equilibrium case corresponding to $\epsilon_i=61.4$ GeV/fm$^3$
is $T_i^{eq}=0.429$ GeV. Therefore there is an interplay in the photon production
between the enhanced temperature and the reduced fugacities in the non-equilibrium.
As a result the total photon yield is suppressed at $p_T=1$ GeV by a factor of five 
but almost identical at $p_T=5$ GeV in the equilibrating compared to the equilibrated
scenario. Keeping, however, the initial temperature ($T_i=0.668$ GeV) instead of the
initial energy density fixed leads to a strong reduction of the non-equilibrium compared to 
the equilibrium rate by one (low $p_T$) to three (high $p_T$) orders of magnitude.
The strong suppression at high $p_T$ is due to the small fugacities in the early hot stage,
from which the energetic photons are emitted mainly. Keeping, however, the initial 
energy density fixed this effect is counterbalanced by the higher initial temperature
in the non-equilibrated case.

The contribution from different physical processes discussed in the preceding section
can easily be identified from Fig.1 itself.
As expected from the equilibrium static rates, the photon spectrum from the equilibrated
QGP (see upper panel of Fig.1) is dominated by the annihilation with scattering contribution. 
In the case of a chemically equilibrating plasma (lower panel of Fig.1) the most dominant 
contribution comes from the usual bremsstrahlung processes, even though the {\it aws} static 
equilibrium rate was higher by almost an order of magnitude. The reason is
obvious as the rate for the {\it aws} processes for an equilibrating plasma involves     
the parton fugacities to cubic power~(Eq.(\ref{awsc})) whereas the other processes are only 
quadratic in the fugacities~(Eqs.(\ref{comp},\ref{ann},\ref{brem2lc})), and the fugacities 
at RHIC are small for the entire life time of the plasma.
At higher $p_T$ the Compton and annihilation one-loop contributions 
even exceed clearly the {\it aws} contribution as energetic photons have their origin in the early
hot stage of the plasma, where the fugacities are very small.
It is also worthwhile to note that the SSPC 
model predicts a gluon dominated plasma implying larger contributions
from processes involving gluons in the entrance channel. 

The total thermal
photon yield at RHIC dominates over the prompt photon contributions \cite{Cle95} for 
$p_T\leq 4.5$ GeV. However, if only the Compton and annihilation contributions 
are considered \cite{Sri97}
the prompt photons overshadow the thermal photon yield already beyond $p_T\geq 3$ GeV.
Since the life time of the plasma at RHIC is small, the photon yield is not affected 
by flow \cite{Sri97}.
   
The corresponding results for LHC energies are shown in Fig.2. Here the non-equilibrium rate
is suppressed compared to the equilibrium at the same initial energy density ($T_i^{eq}=0.695$
GeV) by about a factor of three for all photon momenta and by one to two orders of magnitude
in the case of the same initial temperature. The photon yields from
different processes as well as the total yield are much larger than at RHIC as the life time 
of the plasma, likely to be created at LHC energy, is expected to be much larger and the
initial temperature and fugacities are higher. The upper panel
shows the equilibrated scenario, whereas the lower panel applies to the equilibrating one. 
Now the usual bremsstrahlung
and the {\it aws} contribution originating from two-loop calculations are similar 
in the non-equilibrium case, 
whereas the one-loop contribution is smaller at all $p_T$. 
The reduced suppression of the {\it aws} contribution compared to RHIC 
can be understood in the following way. The rate depends linearly on the temperature 
and cubicly on the fugacities. In addition the {\it aws} is also proportional to  
the energy of the photon. Now with the passage of time the system expands, the temperature 
falls and the chemical reactions pushes the system towards equilibrium causing the fugacities 
to increase, though at later time they decrease significantly caused by transverse expansion.
Hence, there is a competition between the space-time evolution of the temperature and the fugacities 
though they are not exactly counterbalanced. Rather, the interplay of these 
two quantities along with the photon energy dependence causes a reduced suppression 
of the {\it aws} contribution compared to RHIC. 

The prompt photon 
productions \cite{Cle95} due to lowest order QCD (Born) and inclusive photons (background photons 
fragmented off high-$p_T$ quark jets) will remain buried under the thermal photon yield 
for all $p_T$-values considered here. 

Comparing the photon spectra obtained with the wounded-nucleon profile we observe that the
photon yield following from the Fermi-like profile is a little bit higher. This enhancement,
which is always less than a factor of two, is caused by the fact that a Fermi-like profile 
associated with a slower cooling implies a higher temperature.
 
In the following  we present our results with HIJING initial conditions. 
If we use the original prediction, $i.e.$, HIJING-I, then
the life time of the QGP phase is very small (less than 2 fm/c) for RHIC 
and 7.5 fm/c for LHC \cite{Sri97}. Also the matter will be very dilute due 
to the very small initial values of the fugacities at RHIC. Furthermore, for LHC only 
the fluid beyond $4$ fm from the centre participates in flow. Now for HIJING-II initial condition, 
the life time 
of the plasma increases substantially both for RHIC ($\sim$ 4 fm/c) and LHC ($\sim$ 12 fm/c).    
Also the initial fugacities differ by one order of magnitude for RHIC, and are reasonably higher
for LHC.  
For LHC the entire fluid will participate in the transverse flow since the life time is 
large and initially the system will approach equilibrium but then be driven away from 
it as soon as the large transverse velocity gradient develops.
Fig.3 exhibits the photon production from the equilibrating plasma with HIJING-I initial 
conditions and a Fermi-like nuclear profile both at RHIC (upper panel) and LHC (lower panel) 
energies.  Because of the low values of the fugacities during the entire life time of the 
plasma \cite{Sri97}, the contribution of the 
different processes as well as the total yield are strongly suppressed compared to the
case using SSPC initial conditions. In particular there is a huge suppression of the {\it aws}
contribution, depending cubicly on the fugacities,  
compared to the other processes at RHIC. 
Also at LHC the strong suppression of the {\it aws} contribution 
at high-$p_T$ is due to the very small
initial fugacities as high-$p_T$ photons are mostly emitted from the initial stages.

Finally the results for an equilibrating plasma with HIJING-II initial conditions are 
given in Fig.4. Since the initial fugacities are an order of magnitude higher for RHIC
(upper panel),
the contributions from each process, particularly from the {\it aws}, are enhanced substantially
compared to HIJING-I. Since the initial 
fugacities are higher compared to SSPC but the initial temperature is smaller, there is a
counterbalance of these two effects, and the total photon production for LHC (lower panel) is almost 
the same as that 
using SSPC initial conditions (lower panel of Fig.2).

\section{ SUMMARY AND CONCLUSION}

We have considered the photon production from a chemically non-equilibrated QGP at RHIC and
LHC energies. As a new aspect we have included bremsstrahlung processes. The photon production 
rate due to these processes has been calculated recently using the HTL
resummation technique by Aurenche et al. \cite{Aur98} in the case of a fully equilibrated QGP.
Instead of repeating this calculation in the non-equilibrium plasma, we estimated the
non-equilibrium photon production rate in the following way: we simply assigned fugacity factors, describing the
deviation of the parton densities from chemical equilibrium, to the partons in the entrance channels
of the matrix elements corresponding to the different processes (annihilation, Compton scattering, 
bremsstrahlung, annihilation with scattering). In the case of the first two 
processes this
approach has to be shown to give quantitatively very similar results as the more elaborate
HTL method extended to chemical non-equilibrium \cite{Bai97}. Moreover, regarding the uncertainties, 
such as the validity of perturbation theory and the initial
conditions in the computation of the photon spectrum,
this simplification is justified.

The photon spectra have been calculated from these rates by using a hydrodynamical calculation,
describing the space-time evolution of the QGP phase of the fireball, where we have taken
into account the transverse expansion of the fireball. The initial conditions for the temperature 
and the fugacities have been taken from microscopic models (SSPC, HIJING).

We found that the photon yield is reduced by a factor one to five depending on
the collision energy and photon momentum compared
to the fully equilibrated plasma, assuming the same initial energy density taken from the SSPC model. 
This moderate suppression is the result of the interplay between the small fugacities and
the increased temperature in the non-equilibrium compared to the equilibrium scenario. 
In general the suppression is more pronounced  
for energetic photons coming from the highly dilute early stage.  

Assuming the same initial temperature the photon yield from the non-equilibrium QGP is  
reduced by one to three orders of magnitude. The reason for this large difference is
the reduction of the initial temperature for the equilibrated scenario assuming a fixed
initial energy density. In ultrarelativistic heavy ion collisions assuming free flow the 
initial energy density is determined from the measured particle multiplicity,
which is related to the collision energy\cite{Sat92}. Therefore the assumption of a fixed 
initial energy density instead of temperature appears to be more physical.

Although the total photon yield does not differ much in the both scenarios, using the same
initial energy density, its composition by the individual processes is completely different.
Whereas the contribution from the annihilation with scattering process dominates 
the photon production over the entire momentum range in equilibrium, in a chemically 
non-equilibrated plasma it is suppressed at RHIC energies
compared to the bremsstrahlung contribution, which is now dominating, and even
to the one-loop (annihilation, Compton scattering) contributions, which is the smallest
in equilibrium. The reason for this behavior is that the annihilation with scattering
process depends cubicly on the fugacities, which are very small at RHIC, whereas
the other processes only quadratically. 
At LHC energies, on the other hand, where the fugacities are significantly
larger, the annihilation with scattering and the bremsstrahlung contributions are
of the same order and exceed the one-loop contributions clearly. 

We have also investigated the dependence of our results on different initial conditions and
the choice of the profile of the fireball. The photon yield can vary by more than an order
of magnitude, depending on the choice of the initial conditions, while the dependence
on the profile function is weak.

Although in this investigation all important effects (bremsstrahlung, chemical
non-equilibrium, transverse expansion), known so far, are incorporated, there are still
uncertainties and open questions. For example, the applicability of perturbative methods
and the role of higher order contributions is unknown. Also a consistent treatment of the 
chemical equilibration, the space-time evolution of the fireball, and the non-equilibrium
photon rates within the framework of the kinetic theory would be desirable. 

\vspace*{0.5cm}
\centerline{\bf ACKNOWLEDGMENTS}
\vspace*{0.3cm}
The authors are grateful to H. Zaraket for helpful discussions concerning the extension of the
bremsstrahlung rate to non-equilibrium.

\newpage

\begin{table}
\caption{Initial conditions for the hydrodynamical expansion phase
in central collision of two gold nuclei
at BNL RHIC and CERN LHC energies from SSPC and HIJING models.}
\begin{center}
\begin{tabular}{|l|c|c|c|c|c|}
\hline
& & & &&\\
Energy &
$\tau_i$ & $T_i$ & $\lambda_g^{(i)}$&
$\lambda_q^{(i)}$ & $\epsilon_i$ \\
& & & &&\\
 & (fm/$c$) & (GeV)&-&-& (GeV/fm$^3$) \\
& & & & &\\ \hline
{\bf SSPC} & & & &&\\
&&&&&\\
 RHIC & 0.25 & 0.668 & 0.34 & 0.064 & 61.4  \\
& & & & &\\
 LHC & 0.25 & 1.02 & 0.43 & 0.082 & 425 \\
& & & & &\\
{\bf HIJING} &&&&&\\
&&&&&\\
 RHIC, I & 0.7 & 0.55 & 0.05 & 0.008& 4.0 \\
& & & & &\\
 RHIC, II & 0.7 & 0.40 & 0.53 & 0.083& 11.7\\
& & & & &\\
 LHC, I & 0.5 & 0.82 & 0.124 & 0.02&48.6 \\
& & & & &\\
 LHC, II & 0.5 & 0.72 & 0.761 & 0.118&176 \\
& & & & &\\
\hline
\end{tabular}
\end{center}
\end{table}

\newpage

\begin{figure}
\vspace*{-0.8cm}
\centerline{\psfig{figure=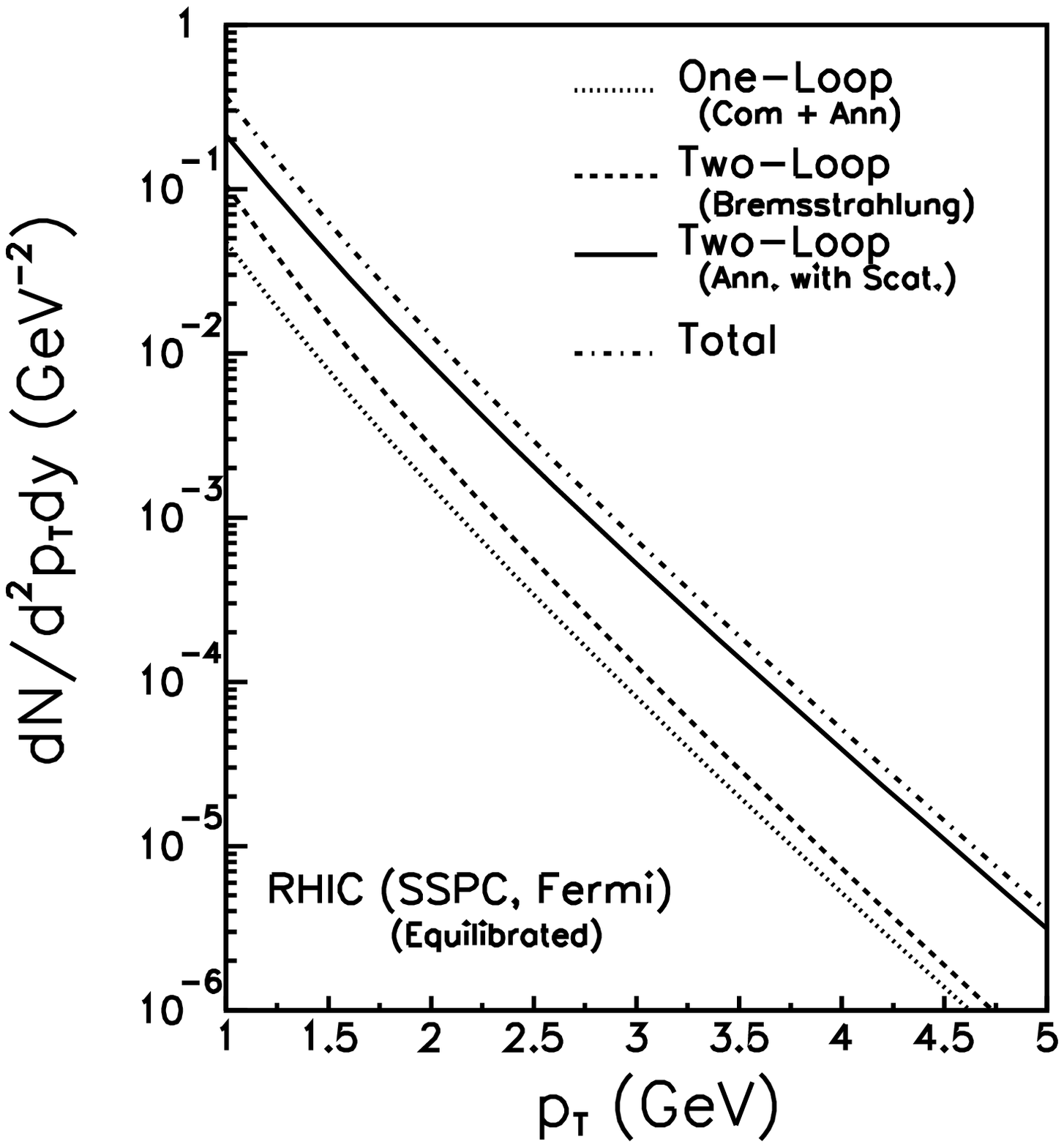,width=14cm,height=11cm}}
\vspace*{-1.9cm}
\centerline{\psfig{figure=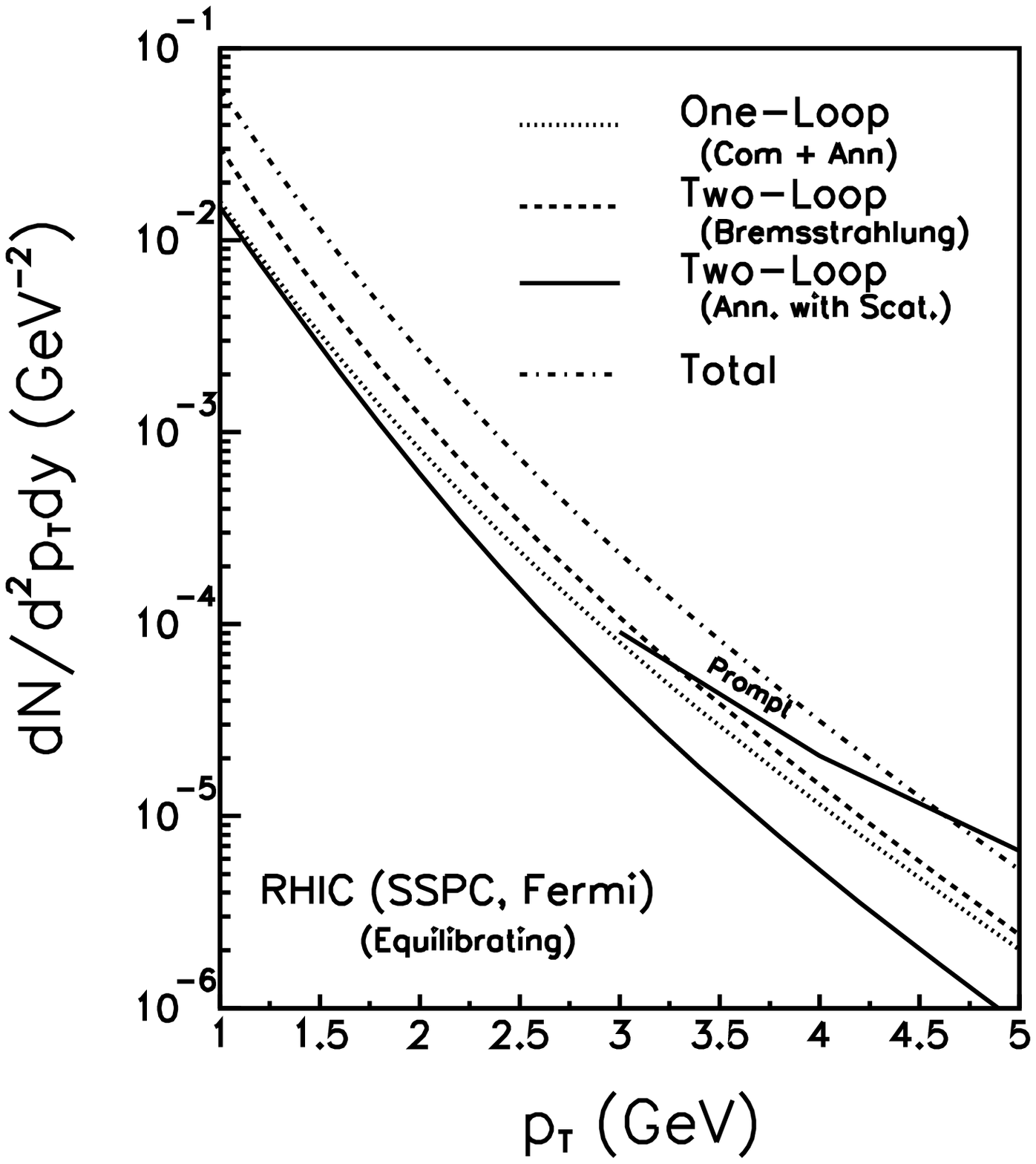,width=14cm,height=11cm}}
\caption{Photon spectra from various processes at RHIC energies 
with SSPC initial conditions and 
the Fermi-like profile function. The upper panel represents the fully 
equilibrated scenario, whereas the lower panel corresponds to the chemically 
equilibrating scenario.}
\end{figure}

\begin{figure}
\vspace*{-0.8cm}
\centerline{\psfig{figure=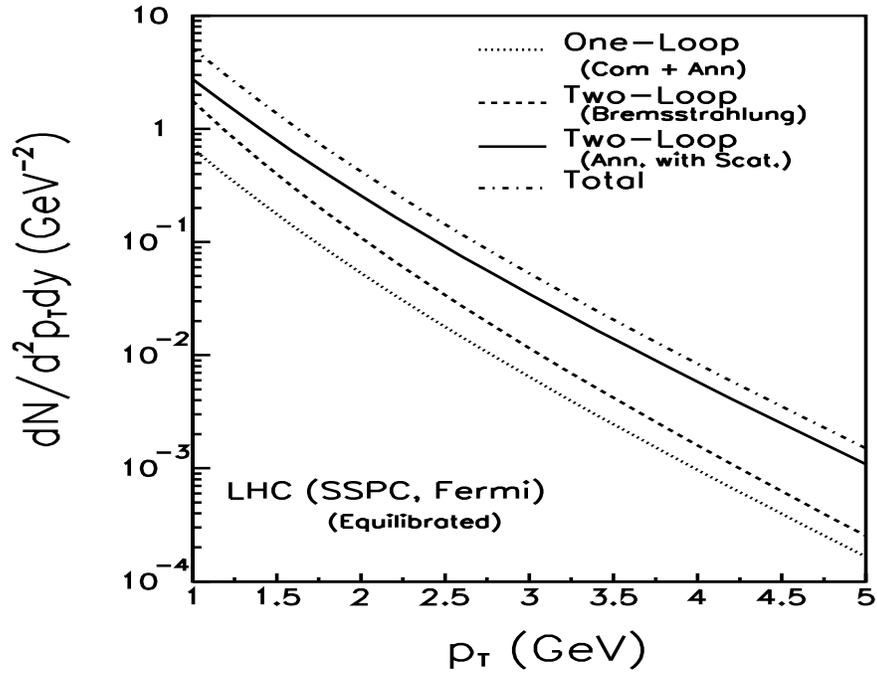,width=14cm,height=12cm}}
\vspace*{-1.9cm}
\centerline{\psfig{figure=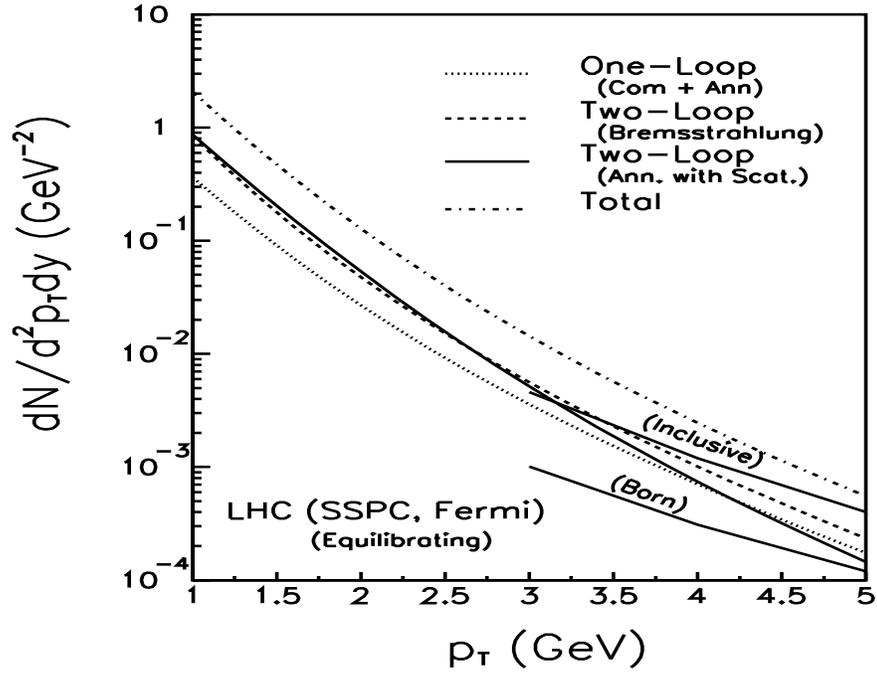,width=14cm,height=12cm}}
\caption{Same as Fig.1 for LHC energies.} 
\end{figure}

\begin{figure}
\vspace*{-0.8cm}
\centerline{\psfig{figure=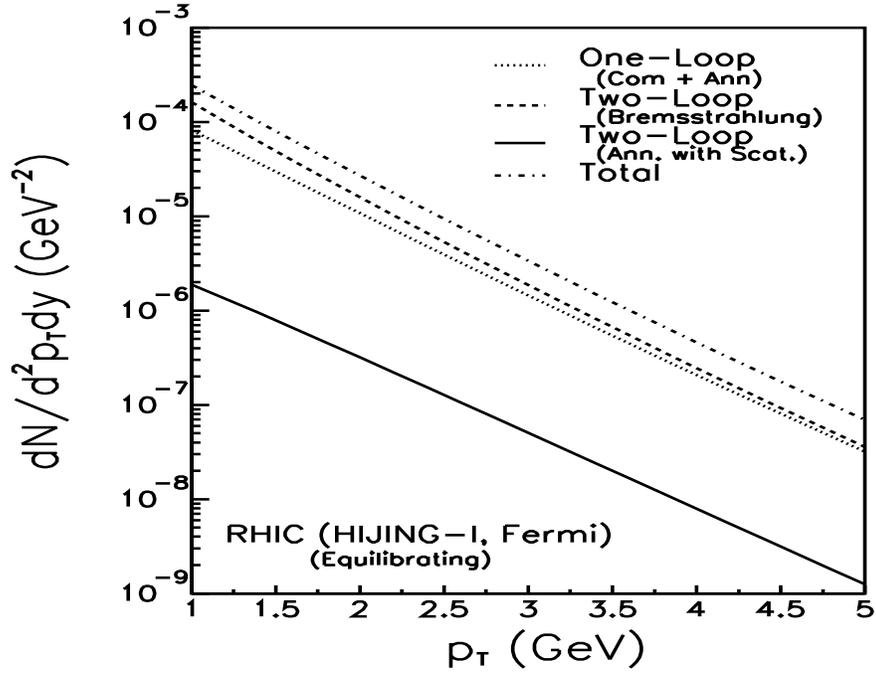,width=14cm,height=12cm}}
\vspace*{-1.9cm}
\centerline{\psfig{figure=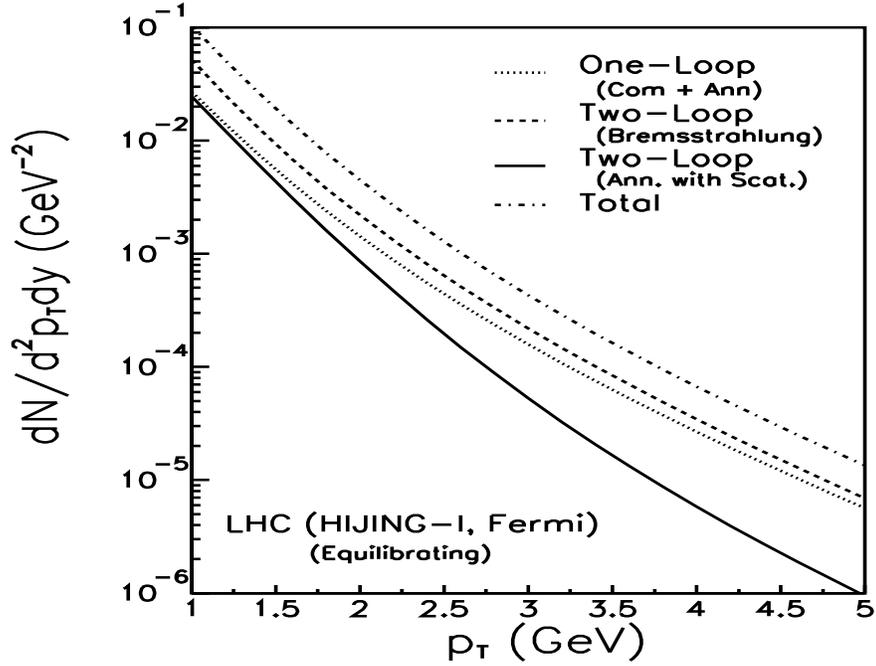,width=14cm,height=12cm}}
\caption{Photon spectra for a chemically equilibrating plasma
at RHIC (upper panel) and LHC (lower panel) energies
 with HIJING-I initial conditions and the Fermi-like profile function.} 
\end{figure}

\begin{figure}
\vspace*{-0.8cm}
\centerline{\psfig{figure=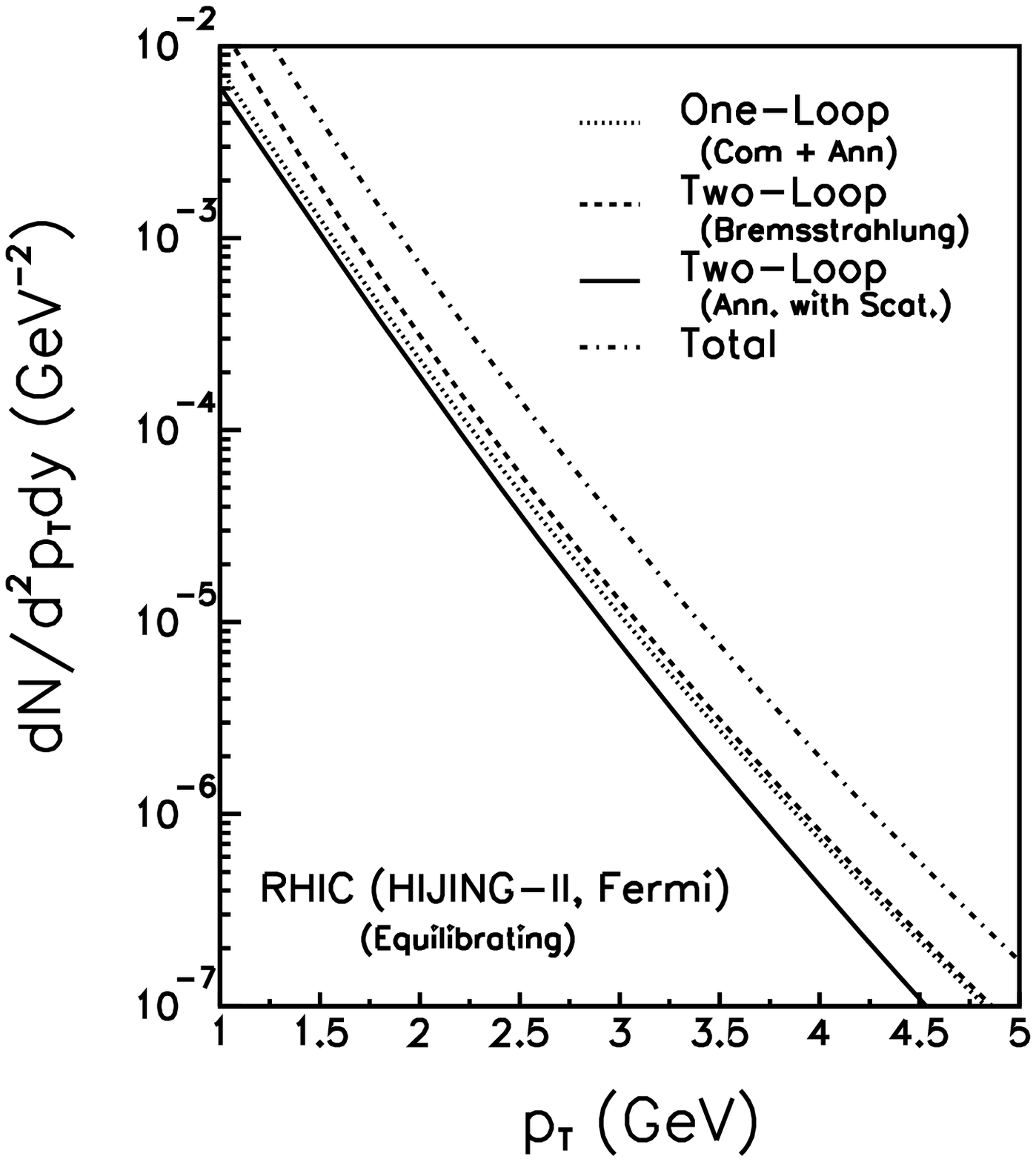,width=14cm,height=12cm}}
\vspace*{-1.9cm}
\centerline{\psfig{figure=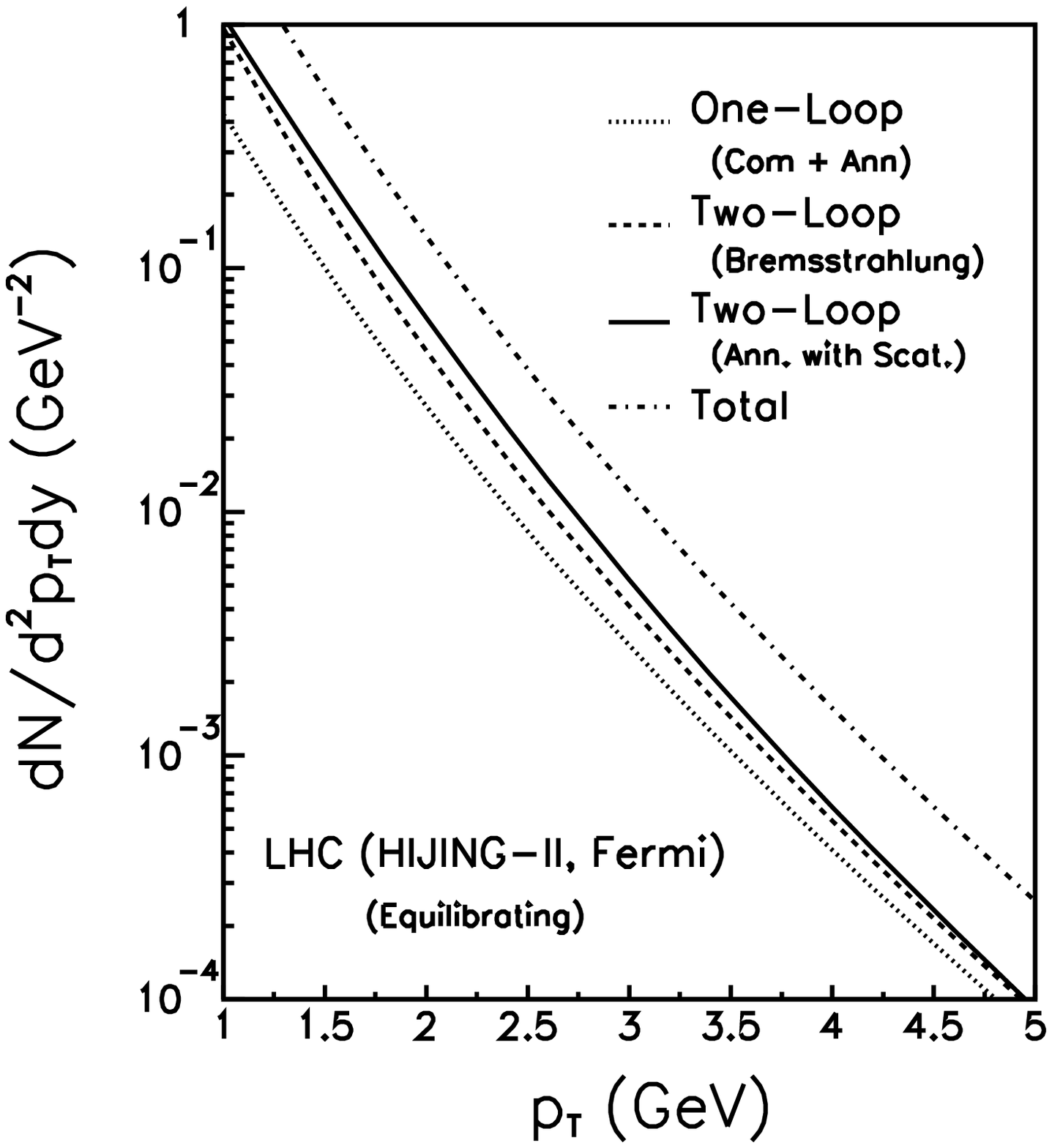,width=14cm,height=12cm}}
\caption{Same as Fig.3 with HIJING-II initial conditions.} 
\end{figure}


\begin{references}
\bibitem{Tho95} M. H. Thoma, Phys. Rev. D {\bf 51}, 862 (1995). 
\bibitem{Ruu92} see for example, P.V. Ruuskanen, Nucl. Phys. A {\bf 544} (1992) 169c.
\bibitem{Aur98} P. Aurenche, F. Gelis, R. Kobes, and H. Zaraket,
Phys. Rev. D {\bf 58}, 085003 (1998). 
\bibitem{Sri99} D. K. Srivastava, Eur. Phys. J. C {\bf 10}, 487 (1999).
D.K Srivastava and B. Sinha, Eur. J. Phys. C {\bf 12}, 109 (2000).
\bibitem{Ste99} F. D. Steffen, $\langle ${\sf nucl-th/9909035}$\rangle $.
\bibitem{Bir93} T. S. Bir\'o, E. van Doorn, B. M\"uller, M. H. Thoma, and X.-N. Wang,
Phys. Rev. C {\bf 48}, 1275 (1993).
\bibitem{Xio94} L. Xiong and E. Shuryak, Phys. Rev. C {\bf 49}, 2203 (1994).
\bibitem{Wan97} X.-N. Wang, Phys. Rep. {\bf 280}, 287 (1997). 
\bibitem{Str94} M. Strickland, Phys. Lett. B {\bf 331}, 245 (1994).
\bibitem{Tra96} C. T. Traxler and M. H. Thoma, Phy. Rev. C {\bf 53},
1348 (1996).
\bibitem{Sri97} D. K. Srivastava, M. G. Mustafa, and B. M\"{u}ller, 
Phys. Rev. C {\bf 56}, 1064 (1997).
\bibitem{Kap91} J. I. Kapusta, P. Lichard, and D. Seibert, Phys. Rev.
D {\bf 44}, 2774 (1991).
\bibitem{Bai92} R. Baier, H. Nakkagawa, A. Niegawa, and K. Redlich,
Z. Phys. C {\bf 53}, 433 (1992).
\bibitem{Sri97a}D. K. Srivastava, M. G. Mustafa, and
B. M\"uller, Phys. Lett. B {\bf 396}, 45 (1997).
\bibitem{Bai97} R. Baier, M. Dirks, K. Redlich and D. Schiff,
Phys. Rev. D {\bf 56}, 2548 (1997).
\bibitem{Car99} M. E. Carrington, H. Defu, and M. H. Thoma, Eur. Phys. J. C {\bf 7} (1999) 347.
\bibitem{Esk96} K. J. Eskola, B. M\"uller, and X.-N. Wang, Phys.
Lett. B {\bf 374}, 20 (1996).
\bibitem{Ger86} H. von Gersdorff, L. McLerran, M. Kataja, and P. V. Ruuskanen,
Phys. Rev. D {\bf 34}, 794 (1986).
\bibitem{Mue97} B. M\"uller, M. G. Mustafa and D. K. Srivastava, Heavy Ion Physics 
{\bf 5}, 387 (1997).
\bibitem{Cle95} J. Cleymans, E. Quack, K. Redlich, and D. K. Srivastava,
Int. J. Mod. Phys. A {\bf 10}, 2941 (1995).
\bibitem{Sat92} H. Satz, Nucl. Phys. A {\bf 544}, 371c (1992).
\end{references}
\end{document}